\documentstyle[12pt,bezier]{article}

\newcommand{\bra}[1]{\langle #1 |}
\newcommand{\ket}[1]{| #1 \rangle}

\newcommand{\braket}[2]{\langle #1 | #2 \rangle}

\newcommand{\bm}[1]{\mbox{\boldmath $#1$}}

\def\const{\mbox{const}}
\def\e{\mbox{e}}

\begin{document}
\title{Modelling macroscopic and baby universes by fundamental strings}
\author{
   V.~A.~Rubakov\\
    {\small \em Institute for Nuclear Research of the Russian Academy of
  Sciences,}\\
  {\small \em 60th October Anniversary prospect 7a, Moscow 117312}\\
  }
\date{May 1995}

\maketitle
\begin{abstract}
We develop a model of
$(1+1)$-dimensional parent and baby
universes  as macroscopic and microscopic fundamental closed strings.
We argue, on the basis of
understanding of strings from the point of view of target $D$-dimensional
space-time, that processes involving baby universes/wormholes not only
induce $c$-number "$\alpha$-parameters" in $(1+1)d$ action, but also lead
to loss of quantum coherence for a $(1+1)d$ observer in the parent
universe.

\end{abstract}

\vskip .5 in

hep-th/9505159

\newpage
\section{Introduction}
It has been argued some time ago
that quantum gravity may allow for processes in which small (say, Planck
size) baby universes branch off the large one, and that these processes
may, among other things, lead to the loss of quantum coherence in the
parent universe \cite{Hawking1,LRT,GS1}. Indeed, a state like
$\ket{\psi}\times \ket{0}_{baby}$ would evolve into a state
\begin{equation}
\ket{\psi'}\times \ket{0}_{baby} + \ket{\psi''}\times \ket{1,I}_{baby}
\label{1*}
\end{equation}
where $\ket{\psi}\dots\ket{\psi''}$ refer to the states of the parent
universe, and $\ket{}_{baby}$ is the state vector in the Hilbert space
of baby universes (i.e.,  $\ket{0}_{baby}$ is the state with no baby
universes, and the state $\ket{1,I}_{baby}$ is the state with one baby
universe of the type $I$).
Since baby universes cannot be probed by an observer in the large
universe, this observer would interprete the state (\ref{1*}) as one
described by a non-trivial density matrix; this would mean
an apparent loss
of quantum coherence in the large universe.

Coleman \cite{Coleman1} and Giddings and Strominger \cite{GS2} put forward
the following argument against this observation.
The effects of baby universes
on  low energy physics in the parent universe may be summarized by
adding extra local terms to the lagrangian
\begin{equation}
    \Delta L(x) = \sum_{I} \hat{A}_{I} O_{I}(x)
\label{2*}
\end{equation}
where $O_{I}$ are local operators composed of  fields living in the
large universe, and $\hat{A}_{I}$ are $x$-independent operators acting
on states of the baby universe subsystem. It has been argued,
furthermore, that the operators $\hat{A}_{I}$ commute with each other;
in an appropriate basis $\hat{A}_{I}=A_{I} + A^{\dagger}_{I}$ where
$A^{\dagger}_{I}$ creates a baby universe of the type $I$ from
$\ket{0}_{baby}$,
and $A_{I}$ is the corresponding annihilation operator. If so, one can
diagonalize the set of operators $\hat{A}_{I}$ by introducing
$\alpha$-states, \[ \hat{A}_{I}\ket{\alpha}=\alpha_{I}\ket{\alpha} \] where
$\alpha_{I}$ are $c$-numbers. These $\alpha$-states are superselection
sectors of the theory, and in a given superselection sector the extra terms
in the lagrangian become \[ \Delta L(x)=\sum_{I}\alpha_{I} O_{I}(x) \] This
means that quantum coherence is restored, and the only effect of baby
universes on the parent universe is the appearance of new coupling
constants $\alpha_{I}$.

The same conclusion has been reached by Klebanov, Susskind and Banks
\cite{KSB} on the basis of the functional integral formalism. However,
further development of this approach has lead Banks \cite{Banks} to the
following picture: the loss of coherence may not be entirely absent in
the closed universe, but suppressed by $\exp (-M_{Pl}^3 V)$ where $V$ is
the volume of the large universe. Even though practically
indistinguishable in a universe like ours, the conclusions of Coleman
and Banks look different in principle; this may be regarded as a signal
that the problem is not completely understood.

A natural model for probing this set of ideas is the theory of
(fundamental) closed strings viewed as the theory of (1+1)-dimensional
universes \cite{Banks2,Hawking2}. It has been realized by Hawking
\cite{Hawking2} and Lyons and Hawking \cite{LyHa} that in this theory,
the $\alpha$-parameters cannot be regarded as $c$-numbers; they should
rather be viewed as  field in mini-superspace, the target space of
strings. In other words, local operators analogous to eq.(\ref{2*}) in
the covariant operator formalism of string theory have the form
\begin{equation}
   A^{\dagger}_{s}(Q)\left(\zeta_{s}^{\mu\nu\dots}
   \partial_{\alpha}X^{\mu}\partial_{\alpha}X^{\nu}\dots\right)\mbox{e}^{-iQX}
    ~+~\mbox{h.c.}
\label{4*}
\end{equation}
(hereafter we consider  strings in critical dimension and
call low lying string states collectively gravitons).
Here $\alpha=0,1$; $\mu,\nu=0\dots,D-1 $,
$D$ is the dimension of the target space-time ($D=26$ for bosonic
string).
$X^{\mu}(\sigma,\tau)$ is
viewed as  field operator in (1+1) dimensions,
while $A^{\dagger}_{s}(Q)$
creates baby universes (gravitons).
Gravitons with different target space momenta $Q$ and different
polarizations $\zeta$ are just different kinds of baby universes; $(Q,s)$
stand for index $I$ in eq.(\ref{2*}); integration over $Q$ and summation
over $s$ is assumed in eq.(\ref{4*}). The observation of Hawking
\cite{Hawking2} corresponds to the fact that operators (\ref{4*}) do not
commute with each other; not only the combination $(A + A^{\dagger})$ but
also $(A - A^{\dagger})$ appears in eq.(\ref{4*}). Coleman's argument
against the loss of quantum coherence apparently does not work.

This observation provides sufficient motivation to take a closer
look, at the tree level of string
interactions,
into the graviton emission by fundamental string interpreted as
branching off of  baby universes in $(1+1)d$ theory.
An advantage of this model is that one can invoke  intuitive
understanding of these processes from the point of view of the
target space (mini-superspace).

There exists fairly strong evidence \cite{Turok1,Turok2,Mitchell}
that heavy fundamental strings with only low harmonics excited (leading
trajectory or alike) behave in flat target space as classical strings
whose length in the c.m. frame is of order of their mass,
\[
    L \sim M
\]
(hereafter $\alpha'$ is set equal to 1/2).
They decay slowly by radiating classical soft gravitational waves with
wavelengths of order $1/L$. These classical strings are long living
objects: the power radiated into gravitational waves is
\[
     \frac{dM}{dt}=\const \times \kappa
\]
where $\kappa$ is the gravitational coupling in $D$ dimensions, and
constant here is of order 1 (actually, it is closer to 100). Thus, it
indeed makes sense to treat these particular string states\footnote{
Clearly, these states are not the most general states of highly excited
strings, but these are the states that may be suitable for modelling large
universes.}
as (1+1)-dimensional universes, even though the necessary formalism
has not yet been elaborated in full detail.

Viewed from (1+1) dimensions, the emission of soft gravitational waves
in target space gives rise to extra terms in the $(1+1)d$ lagrangian,
whose structure
becomes
\[
   \sqrt{g} g^{\alpha\beta} G_{\mu\nu}(X)
    \partial_{\alpha}X^{\mu}\partial_{\beta}X^{\nu}~+~\dots
\]
where  $G_{\mu\nu}$ is the classical target space metrics that includes
the gravitational waves emitted by the string. This is precisely the
picture of classical $\alpha$-parameters\footnote{This
correspondence has been understood by many authors, especially in
the context of $2d$ quantum gravity, see, e.g.,
refs.\cite{Polchinski,Cooper}.}; it is seen to correspond to the
$c$-number approximation of the operators $A^{\dagger}_{s}(Q)$ in the
approximate
description of the emission of gravitons as
the radiation of classical gravitational waves. As
expected \cite{Coleman1,GS2,KSB}, the actual values of the
``$\alpha$-parameters'' $G_{\mu\nu}(X)$ are determined by the history of the
large universe (string); from the target space point of view this is the
history of the radiation of gravitational waves\footnote{But
not only by this history. $G_{\mu\nu}(X)$ are determined also by other
sources of gravity in target space as well as by boundary conditions in
target space (superspace).}.

However, the radiation of classical gravitational waves in target space
is not the whole story. The string (parent universe) may eventually emit
a graviton of relatively high
target space momentum, which will be lost forever for a $(1+1)d$ observer.
This event will be a quantum process, it will not be described by classical
$\alpha$-parameters, and from $(1+1)d$ point of view it will lead to the
loss of quantum coherence. The magnitude of this effect will be
determined by the amplitude of the graviton emission.

In this paper we develop this model of $(1+1)$-dimensional parent and baby
universes -- macroscopic and microscopic strings -- with the main purpose
to discuss the loss of quantum coherence as seen by a $(1+1)d$
observer. The study of microscopic strings in uncompactified
$D$-dimensional space time is technically quite complicated.  We find it
more convenient to consider $D$-dimensional flat space-time with one
spatial dimension, say, $X^{1}$, compactified to a large circle of length
$2\pi L$. Then, as suggested by Polchinski \cite{Polchinski2} and Dai and
Polchinski \cite{DaiPolchinski}, the smooth macroscopic closed string state
$\ket{{\cal P}}$ is naturally constructed as the lowest state that
winds around this compact dimension. For the string at rest, its target
space momentum is (we use the conventions of ref.\cite{GSW})
\[
 {\cal P}=(M_{0},{\bf 0})
\]
where
\[
    M_{0}^{2}=4L^{2} - 8
\]
i.e., the mass is
indeed of order $L$ at large $L$.

In the existing discussions of possible effects of wormholes
\cite{Hawking1,LRT,GS1,Coleman1,GS2,KSB,Banks}, an important role is
played by the interactions of particles, living in the parent
universe, with baby universes that branch off. To model these interactions,
we need particle-like excitations of the fields $X^{\mu}(\sigma,\tau)$ in
$(1+1)d$ universe, in the first place. These are conveniently constructed
by making use of the DDF operators \cite{DDF} that create physical excited
string states by acting on the smooth state $\ket{{\cal P}}$. These
operators are characterized by the mode number $n$, and $n/L$ may be
regarded as the bare $(1+1)d$ momentum (equal to bare $(1+1)d$ energy) of a
"particle" (DDF operators automatically create dressed "particles" whose
total $(1+1)d$ momentum and energy are zero,  as it should be in the closed
universe). The interesting regime is $L\to\infty$ with $n/L$ fixed and
finite. We outline the construction of the smooth string state and its DDF
excitations in section 2.

As discussed above, the loss of quantum coherence occurs when collisions of
particles in $(1+1)d$ universe induce the creation of a baby universe,
i.e., when the macroscopic excited string emits real microscopic sting
states into the target space.
In sections 3 and 4 we consider the simplest of these processes,
namely, those in which both initial and final states of the macroscopic
 string contain two "particles". One property of the string theory (in flat
target space) as the theory of $(1+1)d$ universes is that there exists
global quantum number -- target space momentum $P^{\mu}$ -- and that baby
universes (microscopic strings) carry away this quantum number. As
discussed in section 2, the DDF operators carry {\em light-like} target
space momentum $\Delta P^{\mu}$, so the total momentum of the excited
string is ${\cal P}^{\mu} + \Delta P^{\mu}$. If the emitted microscopic
string is massless (graviton or dilaton), the final macroscopic string
state will typically be a DDF excitation above {\em moving} smooth string;
in other words, zero modes of fields $X^{\mu}(\sigma,\tau)$ will be
excited. Indeed, if the final state is  again the DDF excitation above the
smooth string at rest, then the graviton momentum $Q^{\mu}$ is
\begin{equation}
   Q^{\mu} = \Delta P^{\mu} - \Delta P'^{\mu}
\label{8a}
\end{equation}
where $\Delta P'^{\mu}$ is the target space momentum associated with the
final DDF operators. Equation (\ref{8a}) may hold only when
$Q^{\mu}$,
$\Delta P^{\mu}$ and $\Delta P'^{\mu}$ are collinear,
 otherwise the smooth
part of the final string should carry part of the recoil momentum. In the
latter case the emission of a baby universe involves the interaction with
the entire parent one, so one expects that the corresponding amplitude is
suppressed as $L\to\infty$. We will confirm these expectations  by
explicit calculations in sect.3. What is more important, there is no extra
suppression of the emission probability apart from one just discussed.

If the emitted microscopic string is a tachyon, eq.(\ref{8a}) does not
require
$Q^{\mu}$,
$\Delta P^{\mu}$ and $\Delta P'^{\mu}$ to be collinear.
So, one may expect that the corresponding amplitude is non-vanishing at
  large $L$ in a wider region of phase space.
   We  calculate this  amplitude in sect.4 and
see that it is indeed finite in the above kinematics
in the limit $L\to\infty$.  Unfortunately, the case of two-"particle"
excitation of the final macroscopic string, which we consider in this
paper, is not generic, and the region of the phase space where the emission
 amplitude is unsuppressed is of zero measure. We think, however that
our observations indicate that the total emission probability, and
hence the loss of quantum coherence in $(1+1)d$ universe, is
unsuppressed when the conservation of global quantum numbers does not
require involvement of the entire parent universe into the process of
creation of a baby universe.

Section 5 contains concluding remarks.

\section{Macroscopic strings}

Let one of the spatial dimensions of the target space, $X^{1}$, be
compactified to a large circle of length $2\pi L$. We consider bosonic
closed string theory in critical dimension, in the sector with one string
winding once around this compact dimension. In this sector, the operator
$X^{\mu}(\sigma,\tau)$ is decomposed as follows
\[
   X^{\mu}(\sigma,\tau)=
         X^{\mu} + P^{\mu}\tau + 2L^{\mu}\sigma +
        \frac{i}{2}\sum_{k\neq 0}\frac{1}{k}\left(
	\alpha^{\mu}_{k}
         \mbox{e}^{-2ik\sigma_{+}} +
	\tilde{\alpha}^{\mu}_{k}
         \mbox{e}^{-2ik\sigma_{-}}\right)
\]
where
\[
  L^{\mu} = (0,1,0,\dots ,0)
\]
and other notations follow the conventions of ref.\cite{GSW} (in
particular, $\sigma$ belongs to $(0,\pi)$). It is convenient to introduce
left- and right-moving components of $X^{\mu}(\sigma,\tau)$,
\[
   X^{\mu}_{L}(\sigma_{+})=
         \frac{1}{2}X^{\mu} + \frac{1}{2}P^{\mu}_{L}\sigma_{+} +
         \frac{i}{2}\sum_{k\neq 0}\frac{1}{k}\alpha^{\mu}_{k}
         \mbox{e}^{-2ik\sigma_{+}}
\]
\[
   X^{\mu}_{R}(\sigma_{-})=
         \frac{1}{2}X^{\mu} + \frac{1}{2}P^{\mu}_{R}\sigma_{-} +
         \frac{i}{2}\sum_{k\neq 0}\frac{1}{k}\tilde{\alpha}^{\mu}_{k}
         \mbox{e}^{-2ik\sigma_{-}}
\]
where
\[
   P^{\mu}_{L}= P^{\mu} + 2L^{\mu} ~~~~~~
   P^{\mu}_{R}= P^{\mu} - 2L^{\mu}
\]
The Virasoro operators in the sector with one winding string
are\cite{GSW}
\[
   L_{m} = \frac{1}{2}:\sum_{k} \alpha^{\mu}_{k}
	      \alpha^{\mu}_{m-k}:~~~~
     \tilde{L}_{m} = \frac{1}{2}:\sum_{k} \tilde{\alpha}^{\mu}_{k}
	      \tilde{\alpha}^{\mu}_{m-k}:,~~~~m\neq 0
\]
with
\[
    \alpha^{\mu}_{0} = \frac{1}{2} P_{L}^{\mu}~~~~~~
    \tilde{\alpha}^{\mu}_{0} = \frac{1}{2} P_{R}^{\mu}
\]
and
\[
    L_{0} = \frac{1}{8} P_{L}^{2} + \sum_{k>0} \alpha_{-k}^{\mu}
		       \alpha_{k}^{\mu}~~~~~
       \tilde{L}_{0} = \frac{1}{8} P_{R}^{2} +
    \sum_{k>0} \tilde{\alpha}_{-k}^{\mu}
			      \tilde{\alpha}_{k}^{\mu}
\]

The ground state of the string winding around the compact dimension,
$\ket{{\cal P}}$, is the vacuum of oscillators
$\alpha^{\mu}_{k}$,
$\tilde{\alpha}^{\mu}_{k}$. It has ${\cal P}^{1} = 0$ and
\[
    M_{0}^{2}=-{\cal P}_{\mu}{\cal P}^{\mu} = 4L^{2} - 8
\]

Let us construct the DDF operators that create and annihilate excited
physical states of the winding string. From the $(1+1)d$ point of view,
these states are the states of the large universe with particle-like
excitations of the fields $X^{\mu}(\sigma,\tau)$. The construction begins
with choosing a light-like vector
\begin{equation}
       e^{\mu} = (1,{\bm e}) ~~~~~~
       {\bm e}^{2} = 1
 \label{12c}
 \end{equation}
 and a set of
transverse spatial vectors ${\bm \xi}^{\alpha}$, $\alpha=(1,\dots,D-2)$,
\begin{equation}
   {\bm \xi}^{\alpha} {\bm \xi}^{\beta} =
   \delta^{\alpha,\beta}
\label{12b}
\end{equation}
\begin{equation}
    {\bm \xi}^{\alpha} {\bm e} = 0
\label{12a}
\end{equation}
Then the DDF operators
are defined as follows,
\begin{equation}
   a^{\alpha}_{n}=\int\limits_{0}^{\pi}~\frac{d\sigma_{+}}{\pi}
              \exp\left[4in\frac{e_{\mu}
          X^{\mu}_{L}(\sigma_{+})}{e_{\mu}P^{\mu}_{L}}\right]
	     \xi^{\alpha}_{i}\partial_{+}X^{i}_{L}(\sigma_{+})
\label{13a}
\end{equation}
\begin{equation}
   \tilde{a}^{\alpha}_{\tilde{n}}
   =\int\limits_{0}^{\pi}~\frac{d\sigma_{-}}{\pi}
              \exp\left[4i\tilde{n}\frac{e_{\mu}
          X^{\mu}_{R}(\sigma_{-})}{e_{\mu}P^{\mu}_{R}}\right]
	     \xi^{\alpha}_{i}\partial_{-}X^{i}_{R}(\sigma_{-})
\label{13b}
\end{equation}
Note that the properties of $e^{\mu}$ and ${\bm \xi}^{\alpha}$ ensure that
$e_{\mu}X_{L,R}^{\mu}(\sigma_{\pm})$ commute with $(eP_{L,R})$ and with
$\xi^{\alpha}_{i}\partial_{\pm}X^{i}_{L,R}(\sigma_{\pm})$. Note also that
$(eP_{L,R})$,
$e_{\mu}\alpha^{\mu}_{k}$ and
$e_{\mu}\tilde{\alpha}^{\mu}_{k}$
commute with the DDF operators.

Making use of these properties, it is straightforward to check that the
operators (\ref{13a}), (\ref{13b}) obey the usual oscillator commutational
relations,
\[
   [a^{\alpha}_{n},a^{\beta}_{n'}] = \delta^{\alpha\beta} n \delta_{n+n'}
\]
\[
   [a^{\alpha}_{n},\tilde{a}^{\beta}_{\tilde{n}}] = 0
\]
Their commutational relations with the Virasoro operators can be found
after some algebra,
\begin{equation}
   [L_{m},a^{\alpha}_{n}] =
      -\frac{n}{(eP_{L})} a^{\alpha}_{n} (e_{\mu}\alpha^{\mu}_{m}),~~~~
       m\neq 0
\label{13d}
\end{equation}
\begin{equation}
   [\tilde{L}_{m},a^{\alpha}_{n}] =
      \frac{n}{(eP_{L})} a^{\alpha}_{n}
      (e_{\mu}\tilde{\alpha}^{\mu}_{m}),~~~~ m\neq 0
\label{13e}
\end{equation}
\begin{equation}
   [L_{0},a^{\alpha}_{n}] =
		-\frac{n}{2}a^{\alpha}_{n}
\label{13f}
\end{equation}
\begin{equation}
   [\tilde{L}_{0},a^{\alpha}_{n}] =
		\frac{n}{2}\frac{(eP_{R})}{(eP_{L})}a^{\alpha}_{n}
\label{14a}
\end{equation}
The commutational relations for $\tilde{a}^{\alpha}_{n}$ are obtained from
eqs.(\ref{13d}) -- (\ref{14a}) by interchanging
$a\leftrightarrow\tilde{a}$,
$L_{m}\leftrightarrow\tilde{L}_{m}$,
$P_{L}\leftrightarrow P_{R}$,
$n\leftrightarrow\tilde{n}$,
$\alpha_{k}\leftrightarrow\tilde{\alpha}_{k}$.

Even though the DDF and Virasoro operators  do not commute with each other,
the operators $a^{\alpha}_{n}$ and $\tilde{a}^{\alpha}_{\tilde{n}}$ can be
used for constructing physical states of excited string out of the smooth
state $\ket{{\cal P}}$. Indeed, consider a state
\begin{equation}
    \ket{n,\alpha;\tilde{n},\beta} =
	\frac{1}{\sqrt{n\tilde{n}}}a^{\alpha}_{-n}
	\tilde{a}^{\beta}_{-\tilde{n}} \ket{{\cal P}}
\label{14b}
\end{equation}
where we have chosen the normalization factor in such a way that the norm
of this state coincides with the norm of the smooth string state
$\ket{{\cal P}}$ (in $(1+1)$-dimensional language this corresponds to "one
particle per volume $L$" normalization).
Equations (\ref{13d}) and (\ref{13e}), and similar equations for
$\tilde{a}^{\beta}_{\tilde{n}}$ imply
\[
  L_{m}\ket{n,\alpha;\tilde{n},\beta} =~
  \tilde{L}_{m}\ket{n,\alpha;\tilde{n},\beta} =~0, ~~~~~~m>0
\]
The remaining Virasoro constraints,
\[
  (L_{0}-1)\ket{n,\alpha;\tilde{n},\beta} =~
  (\tilde{L}_{0}-1)\ket{n,\alpha;\tilde{n},\beta} =~0
\]
are satisfied provided that
\begin{equation}
  \frac{n}{(e{\cal P}_{L})}=
  \frac{\tilde{n}}{(e{\cal P}_{R})}
\label{14c}
\end{equation}
The latter condition is the only constraint relating the mode numbers to the
light-like vector $e^{\mu}$. This constraint can be rewritten in the
following form,
\begin{equation}
   n-\tilde{n}=\frac{2(eL)}{(e{\cal P})}(n+\tilde{n})
\label{15b}
\end{equation}
where $(eL)=e_{\mu}L^{\mu}=Le^{1}$. Equation (\ref{14c}) also implies that
\begin{equation}
  \frac{n}{(e{\cal P}_{L})}=
  \frac{\tilde{n}}{(e{\cal P}_{R})} =
  \frac{n+\tilde{n}}{2(e{\cal P})}
\label{15a}
\end{equation}

   Provided this constraint is satisfied, the state
   $\ket{n,\alpha;\tilde{n},\beta}$ is the physical state. It can be viewed
as the dressed oscillator state with mode numbers $n$ and $\tilde{n}$. In
$(1+1)$-dimensional language this state can be interpreted as describing
the large universe with one left-moving
"particle" with bare $(1+1)d$ momentum $n/L$ and
and one right-moving "particle" with momentum $(-\tilde{n}/L)$. One can
construct physical states with more "particles" in a similar way.

   The global quantum number -- target space momentum -- carried by these
   excitations can be read out from the commutational relations of the DDF
   operators with $P^{\mu}$,
\[
   [P^{\mu}, a^{\alpha}_{n}] =
       \frac{2n}{(eP_{L})} e^{\mu} a^{\alpha}_{n}
\]
\[
   [P^{\mu}, \tilde{a}^{\alpha}_{\tilde{n}}] =
       \frac{2\tilde{n}}{(eP_{R})} e^{\mu} \tilde{a}^{\alpha}_{\tilde{n}}
\]
These relations mean that the target space momentum carried by the
operators $a^{\alpha}_{n}$ and $\tilde{a}^{\alpha}_{\tilde{n}}$ is
light-like, $\Delta P^{\mu}\propto e^{\mu}$. In particular, the momentum of
the state $\ket{n,\alpha;\tilde{n},\beta}$ is (see eq.(\ref{15a}))
\begin{equation}
    P^{\mu} = {\cal P}^{\mu} - \frac{2(n+\tilde{n})}{(e{\cal P})} e^{\mu}
\label{16a}
\end{equation}
Note that the mass of this excited string state is
\begin{equation}
     M^{2}=-P^{2}=M_{0}^{2} + 4(n+\tilde{n})
\label{16b}
\end{equation}
which confirms the interpretation of this state in terms of
dressed oscillators.

\section{Emission of microscopic strings with excitation of zero modes}
\subsection{Emission of graviton}
As discussed in Introduction, $(1+1)d$ baby universes are modelled by low
lying string states. Let us first consider the emission of massless states
-- gravitons (and dilatons). We are interested in the following amplitudes,
 \begin{equation}
      \kappa\bra{f}
          :\zeta_{\mu\nu}
          \partial_{+}X^{\mu}(0)\partial_{-}X^{\nu}(0)
          \mbox{e}^{-iQX(0)}:
          \ket{i}
\label{18a}
\end{equation}
where initial and final states $\ket{i}$ and $\ket{f}$ are the DDF-excited
states of the large string, $Q^{\mu}$ and $\zeta^{\mu\nu}$ are the graviton
target space momentum and  polarization. In general, the DDF operators
corresponding to the initial and final states may be different: they may be
constructed with the use of different light-like vectors $e^{\mu}$ and
$e'^{\mu}$ and different transverse vectors ${\bm \xi}^{\alpha}$ and
${\bm \xi}'^{\alpha}$. Thus, in general,
\begin{equation}
  \ket{i}=\frac{1}{\sqrt{n\tilde{n}}} a^{\alpha}_{-n}
           \tilde{a}^{\beta}_{-\tilde{n}}\ket{{\cal P}}
\label{18b}
\end{equation}
\begin{equation}
  \ket{f}=\frac{1}{\sqrt{n'\tilde{n}'}} a'^{\alpha'}_{-n'}
           \tilde{a}'^{\beta'}_{-\tilde{n'}}\ket{{\cal P'}}
\label{18c}
\end{equation}
where $a$ and $\tilde{a}$ are given precisely by
eqs.(\ref{13a}), (\ref{13b}), while $a'$ and $\tilde{a}'$ are defined by
the same formulas with the substitution
$e^{\mu}\rightarrow\e'^{\mu}$,
$\xi^{\alpha}_{i}\rightarrow\xi'^{\alpha}_{i}$.

Making use of eq.(\ref{16a}) one writes the momentum conservation
relation,
\begin{equation}
    Q^{\mu} =
     ({\cal P}^{\mu} -
     {\cal P}'^{\mu})
     - \frac{2(n+\tilde{n})}{(e{\cal P})} e^{\mu}
    + \frac{2(n'+\tilde{n}')}{(e'{\cal P'})} e'^{\mu}
 \label{19a}
\end{equation}
Note that it follows from this relation and eq.(\ref{15b}) that $Q^{1}$ is
quantized in units  $1/L$,
\begin{equation}
	Q^{1} = \frac{r}{L},~~~~~r=0,\pm 1,\dots
\label{19ab}
\end{equation}
as it should be for compact $X^{1}$ (${\cal P}$ and ${\cal P}'$ are also
quantized \cite{GSW}).

 In our case of light-like
 $Q^{\mu}$, eq.(\ref{19a}) implies that
${\cal P}={\cal P}'$
only when $Q^{\mu}$,  $e^{\mu}$ and $e'^{\mu}$ are aligned. For other,
non-exceptional $Q^{\mu}$ the smooth part of the final string state carries
non-zero recoil momentum. In $(1+1)$-dimensional language this means that
the emission of a baby universe with non-exceptional global quantum numbers
$Q^{\mu}$ occurs only when spatially homogeneous modes of the field
$X^{\mu}(\sigma,\tau)$ are excited. This process involves the interaction
with the entire parent universe, and we will see shortly that the
corresponding amplitude is suppressed for universes of large size $L$.

We consider the technically simplest case
\[
     e^{\mu}=e'^{\mu}
\]
\[
     {\bm \xi}^{\alpha} = {\bm \xi}'^{\alpha}
\]
This case includes both the situation with recoil,
${\cal P}\neq{\cal P}'$,
and the exceptional situation without recoil, when
\[
    Q^{\mu} =
     -\left[ \frac{2(n+\tilde{n})}{(e{\cal P})}
    - \frac{2(n'+\tilde{n}')}{(e{\cal P'})}\right] e^{\mu}
\]
Furthermore, we take
\begin{equation}
    e^{1}=0
\label{20d}
\end{equation}
 \begin{equation}
    {\cal P}^{1}={\cal P}'^{1} = 0
\label{20e}
\end{equation}
so that
 \begin{equation}
     Q^{1}=0
\label{20b}
\end{equation}
and also assume that
\begin{equation}
     e^{\mu}\zeta_{\mu\nu}=
     e^{\nu}\zeta_{\mu\nu}=0
 \label{20a}
\end{equation}
These restrictions are purely technical; they simplify the calculations
considerably. Note that eq.(\ref{14c}) implies then
\[
    \tilde{n}=n,~~~~~~~~~~~\tilde{n}'=n'
 \]
Finally, we consider the case
\[
     \alpha'\neq \alpha
 \]
 \begin{equation}
   \beta' \neq \beta
\label{20c}
\end{equation}
which, in $(1+1)$-dimensional language, means that the "particles" change
their $SO(D-2)$ global quantum numbers when interacting with the baby
 universe.

 The calculation of the amplitude (\ref{18a}) is then straightforward.
 The integration over zero modes leads to momentum conservation,
 eq.(\ref{19a}), up to normalization factors about which we will have to
 say more later. The non-zero modes give rise to the product of left and
 right factors,
 \begin{equation}
   A=\kappa\zeta_{\mu\nu}
      \xi^{\alpha}_{i}
      \xi^{\alpha'}_{i'}
      \xi^{\beta}_{j}
      \xi^{\beta'}_{j'}
      A^{\mu}_{L,ii'}
      A^{\nu}_{R,jj'}
 \label{21d}
\end{equation}
The calculation of $A_{L}$ is outlined in Appendix. One finds
 \[
  A^{\mu}_{L,ii'}=
    \frac{1}{2nn'}
    \frac{\Gamma\left(n+1-\frac{n}{(e{\cal P})}(eQ)\right)}
         {\Gamma(n)\Gamma\left(2-\frac{n}{(e{\cal P})}(eQ)\right)}
    \frac{\Gamma\left(n'+1+\frac{n'}{(e{\cal P}')}(eQ)\right)}
         {\Gamma(n')\Gamma\left(2+\frac{n'}{(e{\cal P}')}(eQ)\right)}
	 \times
 \]
 \begin{equation}
	 \left(\frac{{\cal P}_{L}^{\mu}
	 + {\cal P}'^{\mu}_{L}}{8\sqrt{nn'}}U^{i}V^{i'}
	 -\delta^{\mu i}\sqrt{\frac{n}{n'}}V^{i'}
	 -\delta^{\mu i'}\sqrt{\frac{n'}{n}}U^{i}\right)
 \label{21a}
\end{equation}
where
 \begin{equation}
    U^{i} =
      \frac{n}{n-\frac{n}{(e{\cal P})}(eQ)}
      \left[1-\frac{n}{(e{\cal P})}(eQ)\right]
      \left[-Q^{i} + {\cal P}^{i}_{L}\frac{(eQ)}{(e{\cal P})}\right]
 \label{21b}
\end{equation}
 \begin{equation}
    V^{i'} =
      \frac{n'}{n'+\frac{n'}{(e{\cal P}')}(eQ)}
      \left[1+\frac{n'}{(e{\cal P}')}
      (eQ)\right]
      \left[Q^{i'} - {\cal P}'^{i'}_{L}\frac{(eQ)}{(e{\cal P}')}\right]
 \label{21c}
\end{equation}
The factor $A^{\nu}_{R,jj'}$ is obtained from these expressions by
substituting
${\cal P}_{L},{\cal P}'_{L}
 \rightarrow
{\cal P}_{R},{\cal P}'_{R}$,
$i,i' \rightarrow j,j'$, $\mu \rightarrow \nu$.

We are interested in the limit of large $L$ and finite $n/L$, $n'/L$ and
$Q^{\mu}$. The initial string is taken to be at rest,
${\cal P}=(M_{0},0,\dots,0)$. In this limit one has
$(e{\cal P})=(e{\cal P}')=-M_{0}=-2L$,
${\cal P}^{i}_{L}=
{\cal P}'^{i}_{L}=2L\delta^{i,1}$, and using Stirling's formula one
obtains
\[
  A^{\mu}_{L,ii'}=
     \frac{1}{2\Gamma\left(2+\frac{n}{2L}(eQ)\right)
           \Gamma\left(2-\frac{n'}{2L}(eQ)\right)}
	 \left(\frac{{\cal P}_{L}^{\mu}
	  + {\cal P}_{L}'^{\mu}}{8\sqrt{nn'}}U^{i}V^{i'}
	  -\delta^{\mu i}\sqrt{\frac{n}{n'}}V^{i'}
	  -\delta^{\mu i'}\sqrt{\frac{n'}{n}}U^{i}\right)
	  \times
\]
\[
	  \exp\left[(eQ)\left(\frac{n}{2L}\ln{n} -
	   \frac{n'}{2L}\ln{n'}\right)\right]
\]
with
 \[
    U^{i} =
      \left[1+\frac{n}{2L}(eQ)\right]
      \left[-Q^{i} - \delta^{i1}(eQ)\right]
 \]
 \[
    V^{i'} =
      \left[1-\frac{n'}{2L}(eQ)\right]
      \left[Q^{i} + \delta^{i1}(eQ)\right]
\]
We see that the amplitude (\ref{21d}) non-trivially depends on the
parameters of the "particles" (their $SO(D-2)$ flavor and $(1+1)d$ bare
energy $n/L$), and that it behaves at $n\sim n'\sim L$ as
\[
  \left(\frac{1}{L}\right)^{-\frac{n-n'}{L}(eQ)}
\]
Note that the target space energy conservation implies in the limit of
large $L$ that
\[
    Q^{0}=M_{i}-M_{f}=\frac{2(n-n')}{L}
\]
(see eq.(\ref{16b})), so that the suppression factor is
 \begin{equation}
  \left(\frac{1}{L}\right)^{-\frac{1}{2}Q^{0}(eQ)}
 \label{23a}
\end{equation}
The amplitude is finite in the limit $L\to\infty$ only when there is no
recoil into zero modes, i.e., when ${\cal P}={\cal P}'$ and $(eQ)=0$;
otherwise $(eQ)=-Q^{0}+{\bm e}{\bm Q}<0$, and the amplitude vanishes.
This confirms the expectations outlined in Introduction and in the
beginning of this section.

Let us finally count the remaining powers of $L$ in the probability of the
graviton emission. Let all states have the normalization appropriate for
compact $X^{1}$, say, for graviton
$\braket{Q'}{Q}\propto\delta_{r,r'}$ with no   $L$-dependent factors
 ($r$ is
defined in eq.(\ref{19ab})).Then the normalization factors for both string
states and graviton give rise to
the factor $L^{-3/2}$ in the amplitude, while
integration over the zero mode $X^{1}$ in eq.(\ref{18a}) produces the
factor $L$. This leaves the factor $L^{-1}$ in the probability. The
energy-dependent factors in the emission probability, $1/E_{i}E_{f}$, give
another factor
$M_{0}^{-2}\sim L^{-2}$.  The density of states of the graviton and final
string produce the factor $L^{2}dQ^{1}d(n'/L)$, so that the emission
 probability is proportional to $L^{-1}d(n'/L)$. This is precisely the
 volume dependence of the probability of scattering of two "particles" in
 $(1+1)$ dimensions with finite "momenta" $n/L$, given that the states of
 these particles are normalized to contain one particle in volume $L$, see
 eq.(\ref{14b}).  We conclude that apart from the factor (\ref{23a}) there
is no further suppression of the probability of scattering of two
"particles" in the large $(1+1)d$ universe with induced creation of a baby
universe.

Since the transition amplitude is unsuppressed at large $L$ only for
exceptional momenta, i.e., only in the zero measure region of phase space,
the emission probability vanishes too fast in the limit $L\to \infty$.
Thus, the process considered in this section  does not lead to the loss of
quantum coherence in the $(1+1)d$ universe of infinite size. As discussed
above, the origin of this suppression is essentially kinematical, and we do
not expect such a suppression in situations when the excitation of zero
modes is not required by kinematics.
We support this expectation further in section 4 by considering the
emission of a tachyon without recoil into zero modes. Before doing so, let
us briefly discuss the amplitude of the tachyon emission in the case when
the zero modes {\em are} excited.  It will be instructive to see that
tachyons behave qualitatively in the same way as gravitons in this case.

\subsection{Emission of tachyon with recoil into zero modes}

Let us consider the amplitude of the emission of a tachyon with target space
momentum $Q$,
\begin{equation}
      \kappa\bra{f}
          :\mbox{e}^{-iQX(0)}:
          \ket{i}
\label{26a}
\end{equation}
where the states $\ket{i}$ and $\ket{f}$ are defined by eqs.(\ref{18b})
and (\ref{18c}). In this section we again study the particular case
\[
     e^{\mu}=e'^{\mu}
\]
\[
     {\bm \xi}^{\alpha} = {\bm \xi}'^{\alpha}
\]
The target space momentum conservation, eq.(\ref{19a}), implies that in
this case the zero modes are necessarily excited,
\[
     {\cal P}^{\mu} \neq {\cal P'}^{\mu}
\]
We again impose our restrictions (\ref{20d}), (\ref{20e}) and (\ref{20c})
to simplify the calculations.

The evaluation of the tachyon amplitude  (\ref{26a}) is similar (and
simpler) than that outlined in Appendix. One finds, again up to
normalization factors due to zero modes,
 \begin{equation}
   A=\kappa
      \xi^{\alpha}_{i}
      \xi^{\alpha'}_{i'}
      \xi^{\beta}_{j}
      \xi^{\beta'}_{j'}
      A_{L,ii'}
      A_{R,jj'}
 \label{26b}
\end{equation}
where
\[
  A_{L,ii'}=
   - \frac{1}{4\sqrt{nn'}}
    \frac{\Gamma\left(n-\frac{n}{(e{\cal P})}(eQ)\right)}
         {\Gamma(n)\Gamma\left(1-\frac{n}{(e{\cal P})}(eQ)\right)}
    \frac{\Gamma\left(n'+\frac{n'}{(e{\cal P}')}(eQ)\right)}
         {\Gamma(n')\Gamma\left(1+\frac{n'}{(e{\cal P}')}(eQ)\right)}
	 \times
\]
  \[
	 \left(Q^{i} - \frac{(eQ)}{(e{\cal P})}{\cal P}^{i}_{L} \right)
	 \left(Q^{i'} - \frac{(eQ)}{(e{\cal P}')}{\cal P}'^{i'}_{L} \right)
\]
In the intersting limit of large $L$ and finite $n/L$, $n'/L$ and $Q$ (and
the initial string at rest) this expression has the following asymptotics,
\newpage
\[
  A_{L,ii'}=
   - \frac{1}{4\Gamma\left(1+\frac{n}{2L}(eQ)\right)
           \Gamma\left(1-\frac{n'}{2L}(eQ)\right)}
	    \left[Q^{i} + \delta^{i1}(eQ)\right]
	    \left[Q^{i'} + \delta^{i'1}(eQ)\right]
	    \times
\]
\[
           \exp\left[\left(-\frac{1}{2}+\frac{n}{2L}(eQ)\right)\ln{n} +
           \left(-\frac{1}{2}-\frac{n'}{2L}(eQ)\right)\ln{n'}\right]
 \]
 Therefore, the amplitude is of order
 \begin{equation}
   A\sim
  \left(\frac{1}{L}\right)^{2-\frac{n-n'}{L}(eQ)} \sim
   \left(\frac{1}{L}\right)^{2-\frac{1}{2}Q^{0}(eQ)}
\label{27a}
\end{equation}
Given that $Q^{2}=8$ for tachyon, the exponent in this expression is always
positive, and tends to zero at large $|{\bm Q}|$ and
${\bm Q}\propto {\bm e}$. Precisely in this regime the recoil momentum
$({\cal P}' - {\cal P})$ tends to zero. We again find that the amplitude is
unsuppressed only for exceptional tachyon momenta, when the zero modes of
the large string are not excited.

To conclude this section, we point out that the suppression factors like
(\ref{23a}) or (\ref{27a}) are not entirely new in string theory. Similar
suppression appears in the form-factor of the leading trajectory  of large
mass $M\sim L$ \cite{Mitchell}. This suppression should be generic for
macroscopic strings and should allow for the interpretation as coming from
the interaction of  graviton or tachyon with the entire string.

\section{Emission of tachyon without recoil into zero modes}

When the conservation of the target space momentum in the process of
emission of a baby universe does not require the excitation of zero
modes of the fields $X^{\mu}(\sigma,\tau)$ in the parent one, one expects
no suppression of the corresponding amplitude at large $L$. In flat target
space-time this is  possible when the emitted microscopic string state
is a tachyon. So, let us consider the amplitude
 \begin{equation}
 \bra{{\cal P}}a'^{\alpha'}_{n'}\tilde{a}'^{\beta'}_{\tilde{n}'}
      :\mbox{e}^{-iQX(0)}:
       a^{\alpha}_{-n}\tilde{a}^{\beta}_{-\tilde{n}} \ket{{\cal P}}
 \label{29a}
\end{equation}
where ${\cal P}=(M_{0},0,\dots,0)$ for both  initial and final
macroscopic strings, and the DDF operators relevant to the initial and
final states are constructed with different sets of vectors
$(e^{\mu},{\bm \xi}^{\alpha})$ and
$(e'^{\mu},{\bm \xi}'^{\alpha})$, each obeying the relations (\ref{12c}),
(\ref{12b}), (\ref{12a}), i.e.,
 \begin{equation}
 e^{0}=e'^{0}=1
\label{29aa}
\end{equation}
 \begin{equation}
 e^{2}_{\mu}=e'^{2}_{\mu}=0
\label{29b}
\end{equation}
\[
   {\bm \xi}^{\alpha} {\bm \xi}^{\beta} =
   {\bm \xi}'^{\alpha} {\bm \xi}'^{\beta} =
    \delta^{\alpha\beta}
\]
\begin{equation}
    {\bm \xi}^{\alpha} {\bm e} =
    {\bm \xi}'^{\alpha} {\bm e'} = 0
 \label{29c}
 \end{equation}
The target space momentum conservation, eq.(\ref{19a}), in this case reads
\begin{equation}
    Q^{\mu} =
  \frac{2(n+\tilde{n})}{M_{0}} e^{\mu}
    - \frac{2(n'+\tilde{n}')}{M_{0}} e'^{\mu}
 \label{30b}
\end{equation}

We evaluate the amplitude (\ref{29a}) in the general case, without imposing
any restrictions like eqs.(\ref{20d}), (\ref{20e}) or (\ref{20c}). The
oscillator operator algebra involved in the calculation is similar to that
outlined in Appendix. One finds, again up to normalization factors coming
from zero modes,
 \begin{equation}
   A=\kappa
      \xi^{\alpha}_{i}
      \xi^{\alpha'}_{i'}
      \xi^{\beta}_{j}
      \xi^{\beta'}_{j'}
      A_{L,ii'}
      A_{R,jj'}
 \label{30c}
\end{equation}
where
\[
      A_{L,ii'}=
	 \frac{1}{\sqrt{nn'}}\int~\frac{dz}{2\pi}\frac{dz'}{2\pi}
        \frac{1}{z^{n+1}(z')^{n'+1}}
        \left(1-z\right)^{\frac{n}{(e{\cal P}_{L})}(eQ)}
	\times
\]
\begin{equation}
      \left(1-z'\right)^{-\frac{n'}{(e'{\cal P}_{L})}(e'Q)}
     \left(1-zz'\right)^{-\frac{4nn'}{(e{\cal P}_{L})(e'{\cal
     P}_{L})}(ee')} \left[U^{i}(z)V^{i'}(z') +
        \delta^{ii'}\frac{zz'}{(1-zz')^{2}}\right]
 \label{30a}
\end{equation}
where the integration contours are small circles
around the origin
 in the complex plane. Here
 \[
 U^{i}(z)=
    \frac{{\cal P}^{i}_{L}}{2} +
      \frac{Q^{i}}{2}\frac{z}{1-z} -
      \frac{2n'}{(e'{\cal P}_{L})}e'^{i}\frac{zz'}{1-zz'}
 \]
 \[
 V^{i'}(z')=
      \frac{{\cal P}^{i'}_{L}}{2} -
      \frac{Q^{i'}}{2}\frac{z'}{1-z'} -
      \frac{2n}{(e{\cal P}_{L})}e^{i'}\frac{zz'}{1-zz'}
      \]
 We now recall the relations (\ref{15a}),
(\ref{29a}), (\ref{29b}) and (\ref{29c}) and also use $(e{\cal P})=-M_{0}$.
The target space momentum conservation with $Q^2=8$ gives
\[
 \frac{4nn'}{(e{\cal P}_{L})(e'{\cal P}_{L})}(ee')=-1
 \]
 \[
 \frac{n}{(e{\cal P}_{L})}(eQ)=-1
 \]
 \[
 \frac{n'}{(e'{\cal P}_{L})}(e'Q)=1
 \]
These relations make the integration in eq.(\ref{30a}) particularly simple.
We obtain for $n'<n$
 \begin{equation}
      A_{L,ii'}=
      \left(e^{i'}e'^{i}\frac{(n+\tilde{n})}{M_{0}}
      \frac{(n'+\tilde{n'})}{M_{0}} + \delta^{ii'}\right)
      \sqrt{\frac{n'}{n}}
 \label{31a}
\end{equation}
Equivalently,  $A_{L,ii'}$ can be written in the following form (we again
use eqs.(\ref{29b}), (\ref{29c}) and (\ref{30b}))
\[
      A_{L,ii'}=
      \left(\frac{1}{4}Q^{i}Q^{i'} + \delta^{ii'}\right)
      \sqrt{\frac{n'}{n}}
\]
The right factor is obtained in a similar way,
\[
      A_{R,jj'}=
      \left(\frac{1}{4}Q^{j}Q^{j'} + \delta^{jj'}\right)
      \sqrt{\frac{n'}{n}}
\]
We see that the amplitude (\ref{30c}) is finite as $L\to\infty$.

The remaining $L$-dependent factors in the emission probability are counted
in the same way as in the end of sect. 3.1, with the same conclusion.
It can be seen, however, that the final states of the type considered in
this section do not span entire phase space, i.e., the emission amplitude
is unsuppressed only in a zero measure region of the phase space. This
means that at the level of two-"particle" DDF excitations of the final
macroscopic string, the emission probability is still suppressed at large
$L$. We think, however, that our results indicate that the {\em total}
emission probability is finite at large $L$, i.e., the loss of quantum
coherence is likely to occur at finite rate in $(1+1)$-dimensional
universe of large size.

\section{Conclusion}

We have found in this paper that string theory viewed as the theory of
$(1+1)d$  universes meets the expectations on the emission of baby
universes due to interactions of particles in the parent universe. We have
argued, on the basis of
understanding of strings from $D$-dimensional point
of view, that processes involving baby universes/wormholes not only induce
$c$-number $\alpha$-parameters in the $(1+1)d$ action, but also lead to the
loss of quantum coherence for $(1+1)d$ observer in the parent universe.

 We have considered strings in {\em flat} $D$-dimensional
space-time, and restricted ourselves to particular final states of the
macroscopic string, with only {\em two} DDF "particles". The consequence of
these limitations was that only tachyons, or gravitons with very
exceptional momenta, can be emitted without recoil into zero modes. In
both cases the recoil was absent only in a zero measure region of phase
space, so we were unable to show that the emission rate, i.e., the rate
of the loss of quantum coherence, is finite for large $(1+1)d$
universes. Also, the necessity to consider tachyons appears
unsatisfactory.
We think
that the latter peculiarity is not too relevant for our purposes:  indeed,
we have seen in sect.3.2 that tachyons and gravitons behave similarly in
our context. One possibility to improve our analysis  would be to study the
{\em total} probability of the
graviton emission by macroscopis (super)strings {\em in non-trivial
$D$-dimensional background fields}, when the kinematical constraints are
not so restrictive.

 It is not obvious that the simple physical picture evident in string
model of  $(1+1)d$ universes can be extrapolated to (3+1) dimensional
case. In string theory, there exists a natural causal structure of
the $D$-dimensional
target space. It is not clear whether such a structure is inherent in
the superspace of $(3+1)d$ theory. However, it is feasible that
 the notion of baby
universes propagating in (mini-)superspace, which was crucial for our
discussion of the loss of quantum coherence, exists also  in $(3+1)d$
theory: some
of the known examples of wormhole solutions in four dimensions
\cite{GS1,RT,Tamvakis,Farhi}, being appropriately
continued from euclidean time,
describe baby universes that branch off and then evolve non-trivially
in their
intrinsic time (either shrink to singularity or expand to large sizes).
These may be candidates for  baby universes travelling in superspace.

Finally, let us point out that  understanding, in the context of
$(1+1)d$ theory, of  processes involving baby universes may be of
interest for the solution of the information problem in black hole
physics, in view of  suggestions that baby universes/wormholes may
become important at the late stages of the black hole evaporation
(for  discussion and references see, e.g., refs.\cite{Page,Andy,PoSt}).

The author is indebted to T.Banks, M.Douglas, A.Kuznetsov, G.Moore,
Kh.Nirov, S.Shatashvili, M.Shaposhnikov, S.Shenker, D.T.Son, P.Tinyakov and
 N.Turok for numerous helpful discussions. The author thanks Rutgers
 University, where part of this work has been done, for hospitality.
 This work was supported in part by INTAS grant 93-1630.

{\bf Appendix. Amplitude of graviton emission}

Apart from the zero mode integral, the amplitude (\ref{18a}) decomposes
into left and right parts, as written in eq.(\ref{21d}).
Making use of the explicit form of the DDF operators and eqs.(\ref{20b})
and (\ref{20a}), the left factor can be written in the following
form,
\newpage
\[
        A^{\mu}_{L,ii'}=\frac{1}{\sqrt{nn'}}
                  \int\limits_{0}^{\pi}~
                  \frac{d\sigma_{+}}{\pi}
                  \frac{d\sigma_{+}'}{\pi}~
                  \bra{0}
		  \mbox{e}^{2in'\sigma'_{+}}
		  \exp \left[-\frac{2n'}{(e{\cal P}')}
		  \sum_{q>0}\frac{1}{q}e_{\lambda}\alpha^{\lambda}_{q}
		  \mbox{e}^{-2iq\sigma'_{+}}\right]
		  \times
\]
\[
		  \left(\frac{1}{2}{\cal P}'^{i'}_{L}
		  + \sum_{k>0}\alpha^{i'}_{k}\mbox{e}^{-2ik\sigma'_{+}}
		  \right)
		  \exp \left[\frac{1}{2}
		  \sum_{k<0}\frac{1}{k}Q_{\lambda}\alpha^{\lambda}_{k}
		  \right]
		  \left(\frac{{\cal P}^{\mu}_{L} + {\cal P}'^{\mu}_{L}}{4} +
		  \sum_{k\neq 0} \alpha^{\mu}_{k}\right)
		  \times
\]
\[
		  \exp \left[\frac{1}{2}
		  \sum_{k>0}\frac{1}{k}Q_{\lambda}\alpha^{\lambda}_{k}
		  \right]
		  \mbox{e}^{-2in\sigma_{+}}
		  \exp \left[\frac{2n}{(e{\cal P})}
		  \sum_{q<0}\frac{1}{q}e_{\lambda}\alpha^{\lambda}_{q}
		  \mbox{e}^{-2iq\sigma_{+}}\right]
		  \left(\frac{1}{2}{\cal P}^{i}_{L}
		  + \sum_{k<0}\alpha^{i}_{k}\mbox{e}^{-2ik\sigma_{+}}
		  \right)
                  \ket{0}
\]
Moving then the first exponential factor to the right with the use of
eqs.(\ref{20a}) and (\ref{12a}), one picks up the factor
\[
		  \exp \left[\frac{n'}{(e{\cal P}')}(eQ)
		  \sum_{q>0}\frac{1}{q}
		  \mbox{e}^{-2iq\sigma'_{+}}\right]=
		  \left(1-\mbox{e}^{-2i\sigma'_{+}}\right)^
		  {-\frac{n'}{(e{\cal P}')}(eQ)}
\]
Similarly, moving the last exponential factor to the left produces a factor
\[
		  \left(1-\mbox{e}^{2i\sigma_{+}}\right)^
		  {\frac{n}{(e{\cal P})}(eQ)}
 \]
 Moving the first $Q$-dependent exponential factor to the left leads to
 adding a term
 \[
    -\frac{1}{2}Q^{i'}\sum_{k>0}\mbox{e}^{-2ik\sigma'_{+}}=
     -\frac{1}{2}Q^{i'}
     \frac{\mbox{e}^{-2i\sigma'_{+}}}{1-\mbox{e}^{-2i\sigma'_{+}}}
 \]
 to $\frac{1}{2}{\cal P}'^{i'}_{L}$ in the corresponding parenthesis.
 Similarly, a term
 \[
     \frac{1}{2}Q^{i}
     \frac{\mbox{e}^{2i\sigma_{+}}}{1-\mbox{e}^{2i\sigma_{+}}}
 \]
 is added to $\frac{1}{2}{\cal P}^{i}_{L}$ after moving the second
 $Q$-dependent exponential factor to the right.
 The remaining matrix element
 \[
                  \bra{0}
		  \left(\frac{1}{2}{\cal P}'^{i'}_{L}
     -\frac{1}{2}Q^{i'}
     \frac{\mbox{e}^{-2i\sigma'_{+}}}{1-\mbox{e}^{-2i\sigma'_{+}}}
      + \sum_{k>0}\alpha^{i'}_{k}\mbox{e}^{-2ik\sigma'_{+}}
		  \right)
		  \left(\frac{{\cal P}^{\mu}_{L} + {\cal P}'^{\mu}_{L}}{4} +
		  \sum_{k\neq 0} \alpha^{\mu}_{k}\right)
		  \times
\]
  \[
		  \left(\frac{1}{2}{\cal P}^{i}_{L} +
     \frac{1}{2}Q^{i}
     \frac{\mbox{e}^{2i\sigma_{+}}}{1-\mbox{e}^{2i\sigma_{+}}}
      + \sum_{k<0}\alpha^{i}_{k}\mbox{e}^{-2ik\sigma_{+}}
		  \right)
                  \ket{0}
\]
is straightforward to evaluate. Collecting all factors, and denoting
\[
         \mbox{e}^{2i\sigma_{+}}=z ~~~~~~~~
         \mbox{e}^{-2i\sigma'_{+}}=z'
\]
one obtains  at $i\neq i'$ (see eq.(\ref{20c}))
\[
        A^{\mu}_{L,ii'}=\frac{1}{\sqrt{nn'}}
                  \int~
                  \frac{dz}{2\pi}~
                  \frac{dz'}{2\pi}~
		  \frac{1}{z^{n+1}}~
		  \frac{1}{(z')^{n'+1}} ~
		  \times
\]
\[
		  \left[
		  \frac{{\cal P}^{\mu}_{L} +
                 {\cal P}'^{\mu}_{L}}{16} U^{i}(z)V^{i'}(z')
	       +\frac{1}{2}\delta^{i'\mu}
	       \frac{z'}{(1-z')^{2}} U^{i}(z)
	       +\frac{1}{2}\delta^{i\mu}
	       \frac{z}{(1-z)^{2}} V^{i'}(z')
	       \right]
\]
where the integration contours are small circles around the origin in
complex plane and
\[
   U^{i}(z)=
		  {\cal P}^{i}_{L}
                +Q^{i}
               \frac{z}{1-z}
\]
\[
  V^{i'}(z') =
	      {\cal P}'^{i'}_{L}
                -Q^{i'}
               \frac{z'}{1-z'}
  \]

The integrals over $z$ and $z'$ factorize and have the form
\[
      \int\frac{dz}{2\pi}\frac{1}{z^{N+1}}\left(1-z\right)^{\alpha}=
      \left(-1\right)^{N}\frac{\Gamma(\alpha+1)}
      {\Gamma(N+1)\Gamma(\alpha - N + 1)}
\]
Equation (\ref{21a}) is then obtained by simple algebra with the use of the
properties of gamma-function.

\end{document}